\documentclass{aastex62}

\submitjournal{ApJ}

\shorttitle{WD kicks cause the eccentricities of MSP-He WD binaries}
\shortauthors{}

\begin{document}

\title{Asymmetrical mass ejection from proto-white dwarfs and the formation of eccentric millisecond pulsar binaries}

\author{Qin Han and Xiang-Dong Li}
\affil{Department of Astronomy, Nanjing University, Nanjing 210023, China;
lixd@nju.edu.cn}

\affil{Key Laboratory of Modern Astronomy and Astrophysics, Nanjing University,
Ministry of Education, Nanjing 210023, China}

\begin{abstract}
Binary millisecond pulsars (MSPs) are believed to have descended from low-mass X-ray binaries (LMXBs), which have experienced substantial mass transfer and tidal circularization. Therefore, they should have very circular orbits. However, the discovery of several eccentric binary MSPs (with eccentricity $e\sim 0.01-0.1$) challenges this standard picture. Three models have been proposed thus far based on accretion-induced collapse of massive white dwarfs (WDs), neutron star-strange star transition, and formation of circumbinary disks. All of them are subject to various uncertainties, and are not entirely consistent with observations. Here we propose an alternative model taking into account the influence of thermonuclear flashes on proto-WDs. We assume that the flashes lead to asymmetrical mass ejection, which imparts a mild kick on the proto-WDs. By simulating orbital changes of binary MSPs with  multiple shell flashes, we show that it is possible to reproduce the observed eccentricities, provided that the kick velocities are around a few kms$^{-1}$.
\end{abstract}
\large{
\section{Introduction} \label{sec:intro}

Millisecond pulsars (MSPs) are radio pulsars with spin periods shorter than 20 ms and period derivative $\dot{P}$ less than $10^{-19}$\,ss$^{-1}$. The majority of them are in binaries. In the standard formation theory, they are neutron stars (NSs) that have been recycled by accretion from their companion stars in the previous low-mass X-ray binary (LMXB) phase \citep{1982Natur.300..728A,1982CSci...51.1096R}. The companion star finally evolves to be a white dwarf (WD) or  a main-sequence (MS) star of very low mass \citep{1991PhR...203....1B}.  For LMXBs with initial orbital periods longer than the so-called bifurcation period \citep{1988A&A...191...57P}, mass transfer causes the orbit to expand, and there exists a relationship between the mass of the companion and the orbital period at the end of the mass transfer \citep[][hereafter TS99]{PK94,1995MNRAS.273..731R,1999A&A...350..928T}. Meanwhile, tidal torques should bring the binary eccentricities down to $e\sim 10^{-7}-10^{-2}$ \citep{1992RSPTA.341...39P}.

However, several eccentric MSPs (eMSPs) were discovered in recent years, including PSR J1903$+$0327, which is a 2.15 ms MSP in a 95 d eccentric (with $e = 0.44$) orbit around a $1.0 M_\sun$ dwarf companion star \citep{2013MNRAS.435.2234B,2011MNRAS.412.2763F}. The other eMSPs (with $e\sim 0.03-0.1$) form a group with similar orbital periods ($P_{\rm orb}\sim 20-30$ d) and (possible) helium WD (He WD) companions of mass $M_{\rm WD}\sim 0.2-0.3M_{\sun}$ \citep{2010NewAR..54...80B,2013MNRAS.435.2234B,2013ApJ...775...51D,2015ApJ...810...85C,2015ApJ...806..140K}. In addition, this group of binaries seem to obey the WD mass - orbital period relation mentioned above. Their locations in the Galaxy rule out a dynamical origin in a dense stellar environment, so a natural presumption arises that some internal process(es) in the binaries following the mass transfer phase have induced the observed eccentricities.

There are three existing models regarding the origin of the eccentricities. Firstly, \citet{2014MNRAS.438L..86F} proposed a model based on the accretion-induced collapse of super-Chanderasekhar mass WDs. If the collapse is delayed because of the rapid rotation of the WD, the instant loss of gravitational energy during the collapse imposes an eccentricity on the detached binary. They predicted a mass range of $1.22-1.31\ M_\sun$ for the MSPs along with small peculiar space velocities of the binaries ($\sim 10\, {\rm km\,s}^{-1}$). But the predicted MSP mass and space velocity seem to be in tension with current observations of some eMSPs \citep[see e.g.,][]{2016ApJ...830...36A}.

Following the similar idea, \citet{2015ApJ...807...41J} suggested that this sudden loss of mass could happen in the transition from a rapidly rotating NS into a strange star after the mass transfer. Their model constrains the masses of the WD companions to be $0.25-0.28\,M_\sun$. The newborn MSP should be massive enough ($\gtrsim 1.8 M_\sun$) to trigger a phase transition, but the mass of the strange star can lie in a wide range. The uncertainties in this model stem from the undetermined physics of compact stars, including the equation of state and how much binding energy is lost during the phase transition.

\citet{2014ApJ...797L..24A} proposed a model from a different point of view, based on the possible formation of a circumbinary (CB) disk after the mass transfer. When the detached companion, a proto-WD\footnote{We use this term for the detached companions formed after the LMXB phase and prior to the WD cooling stage. For a more precise definition of proto-WD, the readers may refer to, for example, \citet{1975MNRAS.171..555W}.}, experiences thermonuclear flashes due to unstable CNO shell burning, the shell expands and re-fills the Roche-lobe (RL). The rate of the resultant mass transfer typically exceeds the Eddington accretion rate of the NS, so part of the transferred matter may leave from the second Lagrangian point to feed a CB disk. Tidal torques exerted by the CB disk on the binary then excite the orbit to become eccentric, if the disk is sufficiently massive and long-lived. This model suffers from uncertainties in the interaction between the CB disk and the inner binary, and the growth in eccentricities is dependent on the assumptions regarding the magnitude of the viscosity in the disk and possible re-accretion from the disk \citep{2016ApJ...830....8R}.

In this paper, we suggest that the shell flashes on the proto-WDs themselves may considerably influence the orbital evolution by driving asymmetrical mass-loss. We describe the link between the flashes of proto-WDs and the possible kicks imparted on the WDs in Sec.~2. We then introduce the details of the kick model and present the calculated results on the related orbital characteristics in Sec.~3. The uncertainties in our model are also discussed. Finally we summarize in Sec.~4.

\section{asymmetrical mass-ejection during shell flashes on proto-WDs and the resultant kicks} \label{sec:WD_flashes}

We describe the conventional picture of shell flashes in MSP-WD binarie in Sec \ref{sec:flash}, and evaluate the orbital changes they may introduce in MSP-WD binaries in Sec \ref{sec:flash_to_kick}.

\subsection{Shell flashes}\label{sec:flash}

There are numerous studies using 1D stellar evolution code to simulate the evolution of LMXBs and the descendent MSP-WD systems. These studies show that the proto-WDs are able to retain a thin shell of hydrogen (H) outside the core. Moreover, it has been shown that proto-WDs within a certain mass range experience shell flashes due to unstable CNO burning  \citep{1998A&A...339..123D}. The mass interval given in the literature varies from $\sim0.2- 0.3\,M_\sun$ to $\sim 0.2-0.4\,M_\sun$ \citep{2000MNRAS.316...84S,2004ApJ...616.1124N,2001MNRAS.324..617A,2013A&A...557A..19A,2014A&A...571L...3I,2016A&A...595A..35I}. In the previous simulation studies, it was implicitly assumed that part of the thermonuclear energy generated in the flashes is consumed during the spherically symmetrical expansion of the shell. The subsequent RL-overflow episode ensues when the size of the shell exceeds the WD's RL and the matter outside the RL is artificially stripped away. This mass-loss process is considered to exert no direct influence on the orbital eccentricity.

However, the mass-loss processes may be much more complicated than the predictions by the 1D stellar simulation studies. And it will be shown that asymmetrical mass-ejection events may considerably influence the binary orbit.

\subsection{Asymmetrical mass ejection}\label{sec:flash_to_kick}

This issue was addressed in a series of studies by Schaefer and his coworkers \citep{2019MNRAS.487.1120S, 2020MNRAS.492.3323S, 2020MNRAS.492.3343S}. By analyzing the archival data of nova systems, \citet{2020MNRAS.492.3343S} showed that five out of the six examined systems experienced decrease in the orbital periods (with $\left| \Delta P_{\rm orb}/P_{\rm orb}\right| \sim 10^{-5}$) after a classical nova outburst. This is contradictory to the conventional understanding that nova systems should experience orbital expansion as a consequence of spherically symmetrical mass loss from the surface of the WDs. The authors therefore investigated the case of asymmetrical mass loss and demonstrated that it can account for the decrease in the orbital periods\footnote{\citet{2019MNRAS.487.1120S} considered the case of forward asymmetry (along the binary orbit) and showed that an asymmetry parameter of $\epsilon \approx 0.38$ with an ejected velocity of $v\approx 1000\ {\rm km\ s^{-1}}$ is needed to account for the orbital changes of the Nova system QZ Aur. They also pointed out that this degree of asymmetry can be easily achieved in Nova systems.}.

Apart from causing the orbital period changes, asymmetrical mass loss can also introduce orbital eccentricities. If part of the nuclear energy of shell flash(es) is used to accelerate and eject matter asymmetrically, a mild kick could be imparted on the proto-WD due to momentum conservation (we refer to this kick as the WD-kick hereafter). Since orbital circularization in detached NS-WD binaries usually takes an extremely long time \citep{2014ApJ...797L..24A}, the kick-induced eccentricity can then be preserved.

Based on this idea, we examine the influence of the WD-kicks on the orbital characteristics of MSP-WD binaries. We quantify the physical processes during a thermonuclear shell flash with a small amount of instantaneous mass-loss $\Delta M$ along with a WD-kick velocity $V_{\rm k}$.
We first show that the energy $E_{\rm nuc}$ generated during a shell flash is sufficient to accelerate at least part of the shell to a high velocity. The amount of H burned during the flash is typically $\sim 10^{-3.5}\,M_\sun$, approximately the mass increase in the He core mass (we use the characteristic values in \citet{2000MNRAS.316...84S} for the physical quantities of proto-WDs). This gives $E_{\rm nuc}\sim6\times10^{48}$ erg. This energy can accelerate the entire shell (of mass $\sim 10^{-2}\,M_\sun$) to a velocity of $5\times10^{3}\,{\rm km s^{-1}}$ and potentially lead to a sufficiently large WD-kick. Therefore the kick velocity is not severely constrained by energy conservation.

We then estimate the WD-kick velocity $V_{\rm k}$ via momentum conservation $M_{\rm WD}V_{\rm k}=\Delta M V_{\rm ej}$, where $V_{\rm ej}$ is the velocity of the ejecta. It is noted that $\Delta M$ should be smaller than the mass ($\sim 5\times 10^{-3}-10^{-2} M_\sun$) of the shell when the flash occurs \citep{2000MNRAS.316...84S,2016A&A...595A..35I}. If mass ejection takes a similar way during thermonuclear flashes on WDs, we can approximate $V_{\rm ej}$ with the velocity of ejecta in nova systems. Observations show that the ejected shells of novae have a characteristic velocity $\sim 10^3\,{\rm km s^{-1}}$ \citep{2020A&ARv..28....3D}, and the maximum velocities of the ejecta in nova systems are $\sim 5\times 10^3\,{\rm km s^{-1}}$ \citep{2008clno.book.....B}, comparable to the escape velocities of the WDs. Assuming that around 10\% of the material in the shell is ejected with the escape velocity in the opposite direction of the resultant WD-kick, and that the rest of the ejecta has a negligible contribution to the WD-kick, we can approximate the maximum kick velocity as follows,
\begin{equation}
V_{\rm k}\simeq 8\,{\rm kms}^{-1}\left(\frac{\Delta M}{10^{-3} M_\sun}\right)\left(\frac{V_{\rm ej}}{\rm 2.5\times 10^3\,km s^{-1}}\right)\left(\frac{0.3 M_\sun}{M_{\rm WD}}\right).
\end{equation}
In Eq.~(1), $\Delta M$ and $V_{\rm ej}$ are also dependent on $M_{\rm WD}$, but the overall change in $V_{\rm k}$ is small. To achieve a larger kick velocity would require a larger fraction of the WD's shell to be ejected in a collimated form.

\section{the orbital properties of the post-kick binary populations}\label{sec:wd_kick}
We describe the influence of the WD-kicks and analyze the properties of the post-kick binary populations in this section.

We consider the binaries consisting of an MSP and a proto-WD. Since their orbits are expected to be circular ($e<10^{-3}$), we set the initial eccentricities to be zero for simplicity. We take the logarithm of the initial binary orbit periods $\log(P_{\rm orb}/{\rm d})$  to be evenly distributed in the range of $[-1, 3]$. The mass $M_{\rm WD}$ of the proto-WD is calculated from $P_{\rm orb}$ by using the $M_{\rm WD}-P_{\rm orb}$ relation with Solar metallicity in TS99, and the mass of the NS is taken to be $M_{\rm NS}=1.4\ M_\sun$. The initial orbital period $P_{\rm orb}$ is, therefore, the only parameter specifying the initial conditions of an individual binary under these assumptions. We draw a sample of $10^4$ such binaries as the initial population.

The orbital changes in our WD-kick model are computed as follows. The changes in the semi-major axis $a$ and the eccentricity $e$ are governed by the following equations,
\begin{equation}\label{eq:a_evolution}
 \frac{a}{a_0}=[\frac{1-(\Delta M/M_0)}{1-\Delta M/M_0-(V_{\rm k}/V_{\rm 0})^2-2(V_{\rm k}/V_{\rm 0})\cos\theta}],
 \end{equation}
and
 \begin{equation}\label{eq:e_evolution}
1-e^2=\left(\frac{M_0}{M_0-\Delta M}\right)\left(\frac{a_0}{a}\right),
\end{equation}
where the quantities with and without subscript $0$ correspond to the pre-kick and post-kick binaries, respectively; $M_0=M_{\rm WD}+M_{\rm NS}$ is the mass of the binary, $V_{\rm 0}=(GM_0/a_0)^{1/2}$ is the orbital velocity, and $\theta$ is the angle between the direction vectors of $V_{\rm k}$ and $V_{\rm 0}$.

We define two dimensionless factors,
\[f_{\rm m}=\frac{\Delta M}{M_0},\] and \[f_{\rm k}=\frac{V_{\rm k}}{V_{\rm 0}}.\]
Using the TS99 relation we can parameterise $f_{\rm k}$ with $V_{\rm k}$ and $M_{\rm WD}$,
\begin{equation}\label{eq:fk_ex}
f_{\rm k}=(\frac{V_{\rm k}}{\rm 4.32\ kms^{-1}})({M_{\rm WD}\over M_\sun}-0.12)^{3/2}.
\end{equation}
Then the eccentricity of the post-kick orbit can be expressed as,
\begin{equation}\label{eq:e_ex}
e=\frac{\sqrt{f_{\rm m}^2-f_{\rm m}+f_{\rm k}^2+2f_{\rm k}\cos\theta}}{1-f_{\rm m}}.
\end{equation}

We use a subroutine in the binary population synthesis (BPS) code developed by \citet{2002MNRAS.329..897H} to compute the orbital evolution.

\subsection{Eccentricities caused by single kick}\label{sec:k1}
We start by imparting one kick on the proto-WDs and investigate the orbital properties of the post-kick population. We tentatively assume that the magnitude of the kick velocity follows a Maxwell distribution specified with a parameter $\sigma _{\rm k}$. We then construct three scenarios with the same amount of (asymmetric) mass-loss $\Delta M=10^{-3}\,M_\sun$, but with different values of $\sigma_{\rm k}$, i.e., $\sigma_{\rm k}=1,\ 2$, and $3\ {\rm kms^{-1}}$. In the following we use the properties of the kick(s) to name the post-kick populations, such as $K_{1}\sigma_{1}$, $K_{1}\sigma_{2}$ and $K_{1}\sigma_{3}$. Here the subscripts of $K$ and $\sigma$ denote the number of kicks and the magnitude of $\sigma$ (in units of kms$^{-1}$), respectively.

The eccentricities of the post-kick systems are plotted against the masses of the WDs in Fig.~\ref{fig:Fig1}. The left, middle, and right panels correspond to populations $K_{1}\sigma_{1}$, $K_{1}\sigma_{2}$ and $K_{1}\sigma_{3}$, respectively.
The colored dots mark the simulated post-kick systems. The magnitude of the kick velocity can be read from the right color bar.
We plot the five eccentric MSPs in green dots with error bars, which cover the $90\%$ probability mass range for randomly oriented orbits (the measured and derived parameters for the five binaries are listed in Table \ref{table:2}).
The black dotted line represents the eccentricity caused by $10^{-3}\ M_\sun$ mass loss without any kick.

From Fig.~1 we can see that the kick can substantially affect the distribution of the eccentricity. While population $K_{1}\sigma_{3}$ covers the parameter space where eMSPs occupy, populations $K_{1}\sigma_{1}$ and $K_{1}\sigma_{2}$ with relatively smaller kick velocities can only reproduce some of the observed eMSPs. In all the three panels, there is a trend that the magnitude of the induced eccentricities increases with the orbital period. This is because with the same $V_{\rm k}$, $f_{\rm k}$ is larger for systems with larger $M_{\rm WD}$ (see Eq.~[\ref{eq:fk_ex}]) or longer $P_{\rm orb}$.

However, we need to point out that, to reproduce the eccentricities of $\sim 0.1$ in the eMSP binaries with $P_{\rm orb}\sim 30$ d, the required kick velocities are $\sim 4.0-10.0\ {\rm kms^{-1}}$. These values are close to or even exceed the maximum achievable kick velocity we calculated above. So it is probably difficult to attribute the origin of the observed eccentricities to single shell flash event.

\subsection{Eccentricities caused by multiple kicks}\label{sec:k_mul}
Multiple flashes have been often observed in the theoretical studies of the evolution of proto-WDs. The majority of proto-WDs are expected to experience less than 5 kicks  \citep{2000MNRAS.316...84S,2004ApJ...616.1124N,2001MNRAS.324..617A,2013A&A...557A..19A}, and the maximum number of kicks is 26 in \citet{2016A&A...595A..35I}. We therefore conduct simulations with multiple kicks to investigate whether they can reproduce the observed eccentricities and to constrain the magnitude of the kick velocity.

We start from one up to 15 kicks, and plot the results with 1, 2, 3, 5, 10, 15 kicks on the $e-M_{\rm WD}$ plane in panels A to E of Fig.~\ref{fig:Fig2}, respectively. To avoid the total ejecta mass getting too large, we set $\Delta M=1\times 10^{-4}\,M_\sun$ and $\sigma_{\rm k}=1\ {\rm kms^{-1}}$ in each flash. The colors of the scattered points mark the magnitude of the latest kick velocity, that is, the 1st, 2nd, 3rd, 5th, 10th and 15th kick velocity for panels A to F, respectively. We can see from Fig.~\ref{fig:Fig2} that the $K_{10}\sigma_{1}$ and $K_{15}\sigma_{1}$ populations cover the parameter space where the observed eMSPs occupy. Therefore, we need approximately at least 10 kicks to reproduce the eMSPs.

In order to compare the influence of the kick numbers in more detail, we plot the number density distributions of populations $K_{1}\sigma_{3}$ and $K_{15}\sigma_{1}$  on the $P_{\rm orb}-e$ plane in the left panels (A and C respectively) of Fig.~\ref{fig:Fig3}.  In both cases we have similar number density distributions. Then, we select binaries with eccentricities larger than a threshold value, $e_{\rm crit}$, and plot the cumulative number distributions in the right panels (B and D respectively) for the two populations (the distributions of binaries from populations $K_{5}\sigma_{1}$ and $K_{10}\sigma_{1}$ are also plotted in panel D). The dashed and solid lines correspond to $e_{\rm crit}=0.01$ and 0.1, respectively. Comparing the lines with $N_{\rm k}=5$, 10, and 15, we can see that the $e$ distributions become more and more similar when $N_{\rm k}$ increases.

To further examine this trend, we display the $e$ distributions of populations $K_{i}\sigma_{1}$ (where $i$ ranges from 1 to 15) in Fig.~\ref{fig:Fig4}. They are plotted with different colors which represent the number of kicks. The vertical lines in the same color as the  distribution curve represents its mean value, and the two black vertical lines represent $e=0.01$ and 0.1. We also see that the distributions become more and more similar from the top panel to the bottom panel. In order to check whether they have (statistically) evolved into a ``saturated" state, we conduct two sample Kolmogorov-Smirnov (KS) and Anderson-Darling (AD) tests of the $e$ distributions for populations $K_{i}\sigma_{1}$ and $K_{i+1}\sigma_{1}$ ($i$ =1 to 14).
The null hypothesis is that the eccentricities of populations $K_{i}\sigma_{1}$ and $K_{i+1}\sigma_{1}$ come from the same distribution function. For the two sample KS test, we choose the critical KS value as $S_{\rm ks\_ 2samp}=0.01$ and the $P$ value as $P_{\rm ks\_ 2samp}=0.01$; and for the AD test, we choose a significance level of $5\%$. Both tests show that the null hypothesis cannot be rejected when the number of kicks is larger than 11, that is, the  $e$ distribution has reached a ``saturated" state after 12 kicks.

We also adopt other values of $\sigma_{\rm k}$ to examine its influence. When $\sigma_{\rm k}=0.75\ {\rm kms^{-1}}$, approximately 10 kicks can still reproduce the $e$ distribution of the eMSPs; when  $\sigma_{\rm k}=0.50\ {\rm kms^{-1}}$, approximately 15 kicks are needed for the observed eMSPs.

\subsection{Comparison with the MSP-He WD binaries }\label{sec:compare_obs}
In this subsection, we compare the simulated post-kick populations with the entire population of the observed MSP-He WD binaries. Since the orbital periods are more accurately determined than the WD masses, we choose to compare our results with the observations on the $P_{\rm orb}-e$ plane.

As shown in Figs.~\ref{fig:Fig1} and \ref{fig:Fig2}, populations $K_{1}\sigma_{3}$ and $K_{15}\sigma_{1}$ contain eMSPs similar to the observed ones, so we first compare them with the observed MSP-He WD binaries in panels A and C of Fig.~\ref{fig:Fig5}, respectively. One can see that the distributions of the post-kick populations clearly deviate from that of the observed MSP-He WD binaries. The small group of eMSPs are the only ones that overlap with our simulated populations. This discrepancy suggests that if the kick scenario applies, a certain mass interval for the WD-kick is required.

We therefore add a mass interval $[M_{\rm min},M_{\rm max}]$ for shell flashes to our model. Currently, there is no consensus on the exact values of $M_{\rm min}$ and $M_{\rm max}$, so we arbitrarily set them to be 0.268 $M_\sun$ and 0.281 $M_\sun$, which are converted from the minimum and maximum $P_{\rm orb}$ (22 and 32 d) of known eMSPs. We then redo the simulation assuming that kick(s) are only imparted on the proto-WDs with masses within that interval, and other proto-WDs only experience spherically ejected mass loss.
The simulated $K_{1}\sigma_{3}$ and $K_{15}\sigma_{1}$ populations are plotted in panels B and D of Fig.~\ref{fig:Fig5}, respectively. Comparing panels B and D with panels A and C, we can see that the mass interval is translated into an ``orbital interval" due to the $M_{\rm WD}-P_{\rm orb}$ relation, producing a distinct group with relatively high eccentricities on the $P_{\rm orb}-e$ plane.
Note that the kick(s) also cause the orbital period to slightly change, so the range of the orbital periods of eMSPs (with $e\gtrsim 10^{-2}$) are somewhat broadened to be $[18,\ 45]$ d and $[19,\ 39]$ d for populations $K_{1}\sigma_{3}$ and $K_{15}\sigma_{1}$, respectively.

Therefore, the mass interval is of great importance for making predictions. However, the conditions for establishing shell flashes are still uncertain, which closely depend on the mass retained in the shell at the moment of detachment and its chemical structure. So it is not surprising that different combinations of the related physical ingredients in the recipes of stellar models, as well as the initial conditions such as metallicity and the WD mass, lead to different boundary values of the mass interval \citep[e.g.][]{2014A&A...571L...3I}.

Our simulated population does not match the observations of normal MSP-He WD binaries. That is not unexpected, since we have adopted an initial zero eccentricity for simplicity while the actual eccentricities of MSP-He WD binaries should be largely determined by the tidal interactions between the NS and its low-mass companion \citep{1992RSPTA.341...39P}. In panel B of Fig.~\ref{fig:Fig5}, the simulated eccentricities are distributed within a strip because there is only one mass loss event with fixed amount of the ejecta mass, while in real situations, the ejecta mass is likely to be diverse and dependent on the mass of the proto-WD. In comparison, the simulated eccentricities are distributed over a wider range in panel D, suggesting that multiple mass-loss events could also be partly responsible for the small eccentricities of normal MSP-He WD binaries.

To investigate the influence of the ejecta mass and the kick in some detail, we conduct simulations taking into account other distribution laws for these two factors. We assume that all proto-WDs experience WD-flashes, and that both the ejecta mass and the magnitude of the kick velocity follow a power-law distribution. We define a factor $f_a$ to quantify the asymmetry in the mass-loss processes during shell flashes, and use this factor to parameterize the kick imparted on the proto-WD. Specifically, we let $\log f_a$ to be randomly chosen from a flat distribution between $-5$ and $0$, the ejecta mass and the kick velocity are then determined with $\log (\Delta M/M_\sun)=-3.0+\log f_a$ and $\log (V_k/{\rm kms}^{-1})=\log(8)+\log f_a$, respectively. The simulated post-kick populations are plotted in Fig.~\ref{fig:Fig6}. The left panel shows the number distribution of the simulated population, and the right panel compares the simulated and observed distributions. We can see that if the degree of asymmetry in the mass loss during a shell flash can vary over several orders of magnitude, the distribution of the induced eccentricities is able to cover the entire eccentricity range of the observed systems.

\section{Summary}\label{sec:summary}
To summarize, we propose a model based on the shell flashes of proto-WDs to explain the abnormal eccentricities of some MSP-He WD binaries. Assuming that thermonuclear shell flashes power asymmetrical mass loss and result in mild kicks on the proto-WDs, we simulate the eccentricity distribution of the post-kick populations under various situations.
Our results show that it is possible to account for the eccentricities of eMSPs, if reasonable choices of mass loss and kick velocities during shell flashes are adopted. On one hand, if the occurrences of shell flashes are distinct for proto-WDs of mass within a specific interval, then we would expect the orbital periods of eMSP binaries to be distributed within a particular range, as the current sample have shown. If, on the other hand, the mass interval is quite large, then we would expect to find eMSP binaries over a wide orbital period range in the future.

\acknowledgments

We are grateful to the referee for valuable comments. We also thank Prof. Kinwah Wu for helpful suggestions. This work was supported by the National Key Research and
Development Program of China (2016YFA0400803), the Natural Science Foundation of China under grant No. 11773015, 10241301, and Project U1838201 supported by NSFC and CAS.

\clearpage
\begin{figure}
	\centering
	\includegraphics[width=1.0\textwidth]{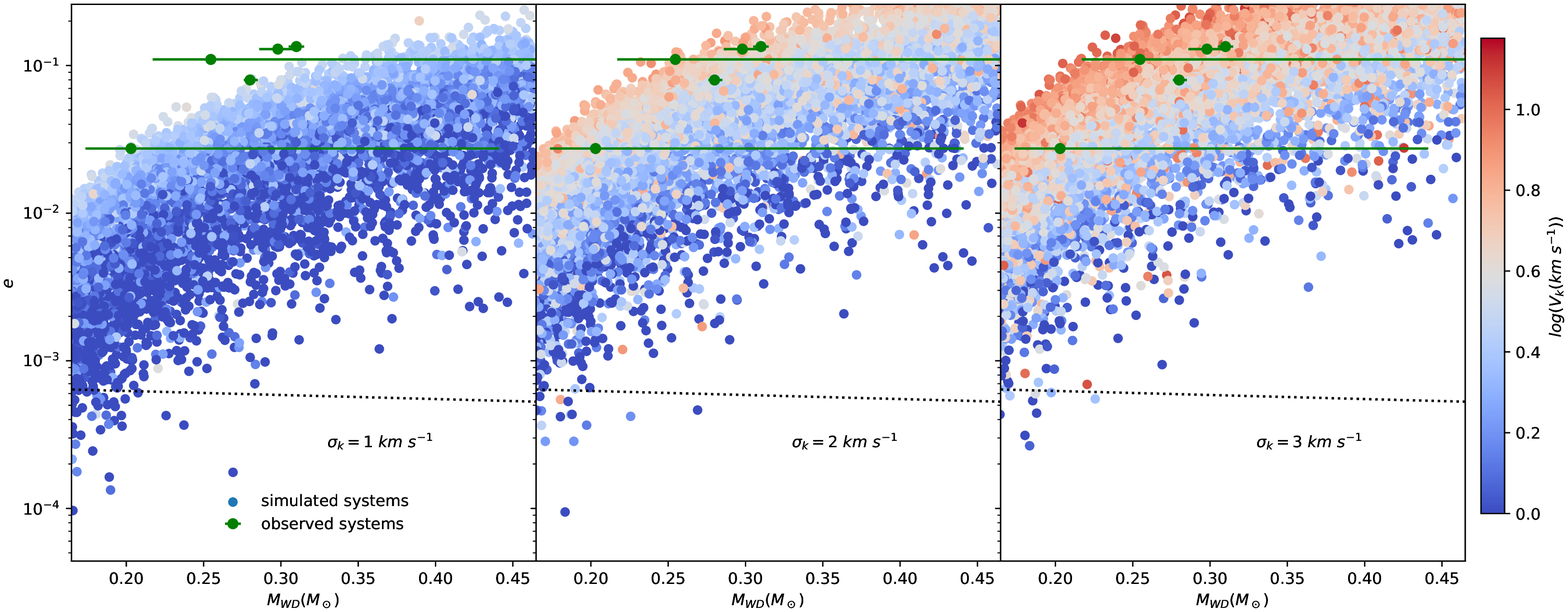}
	\caption{Simulated distributions of MSP-He WD binaries on the $e-M_{\rm WD}$ plane. We assume that the proto-WD was imparted with a single kick  and use different colors to indicate the magnitude of the kick velocity $V_{\rm k}$. The amount of mass loss is taken to be $10^{-3}\,M_\sun$. In the left, middle, and right panels the Maxwellian distribution parameter $\sigma_{\rm k}$ equals to 1, 2, and 3 kms$^{-1}$, respectively. The black dotted line represent the results with no kick considered.
The green dots with error bars represent the five eMSPs. }
\label{fig:Fig1}
\end{figure}

\begin{figure}
	\centering
	\includegraphics[width=1.0\textwidth]{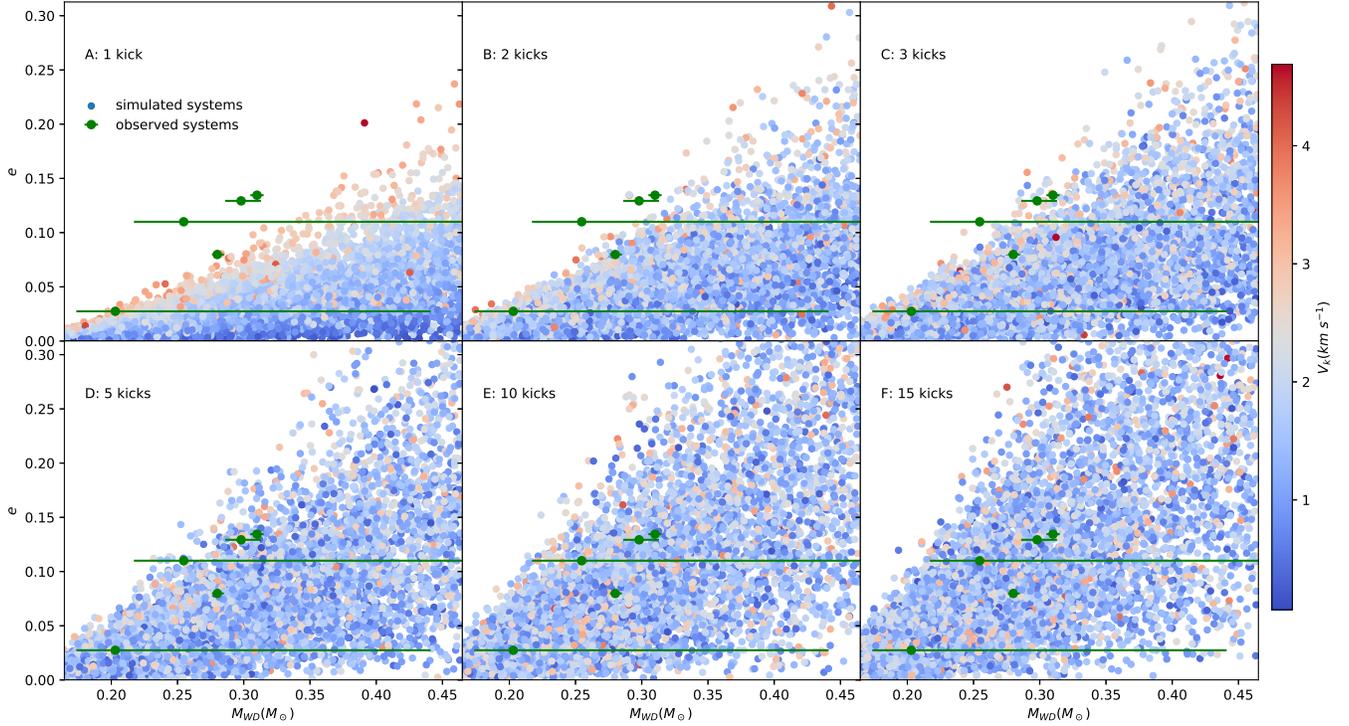}
	\caption{Same as Fig.~1 but for multiple kicks. Panel A - E corresponds to populations that have experienced 1, 2, 3, 5, 10, and 15 kicks with $\sigma_{\rm k}=1\ {\rm kms^{-1}}$, respectively.}\label{fig:Fig2}
\end{figure}

\begin{figure}
	\centering
	\includegraphics[width=0.7\textwidth]{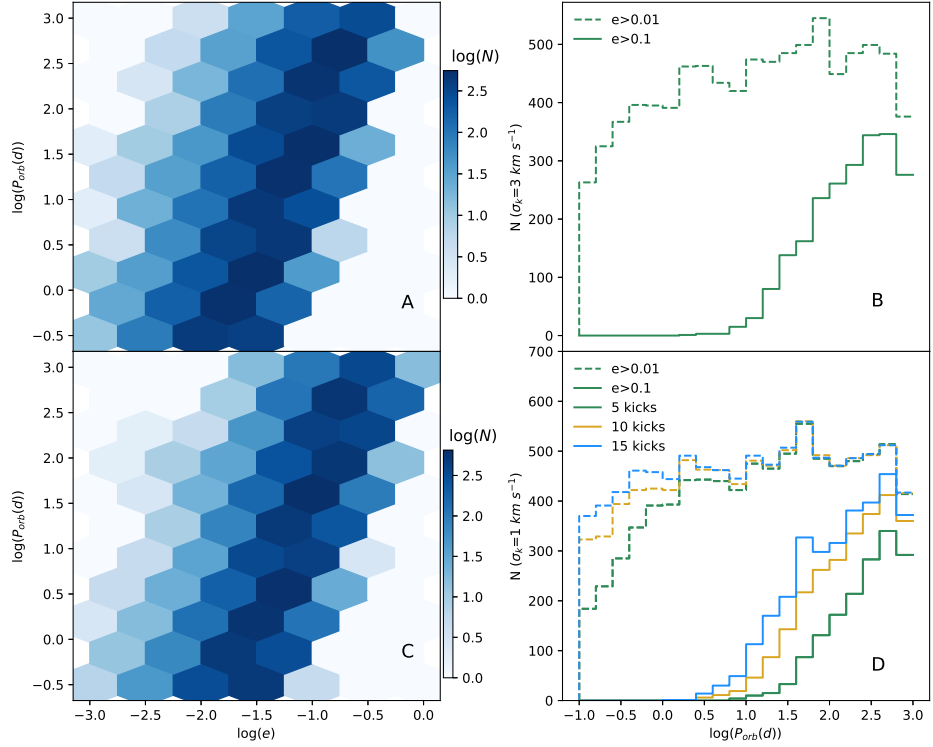}
	\caption{The number distributions of post-kick populations on the $P_{\rm orb}-e$ plane (left panels) and the cumulative number distribution of eMSP binaries (right panel). The upper and lower panels correspond to populations $K_{1}\sigma_{3}$ and $K_{15}\sigma_{1}$, respectively.
	}\label{fig:Fig3}
\end{figure}

 \begin{figure}
	\centering
	\includegraphics[width=0.4\textwidth]{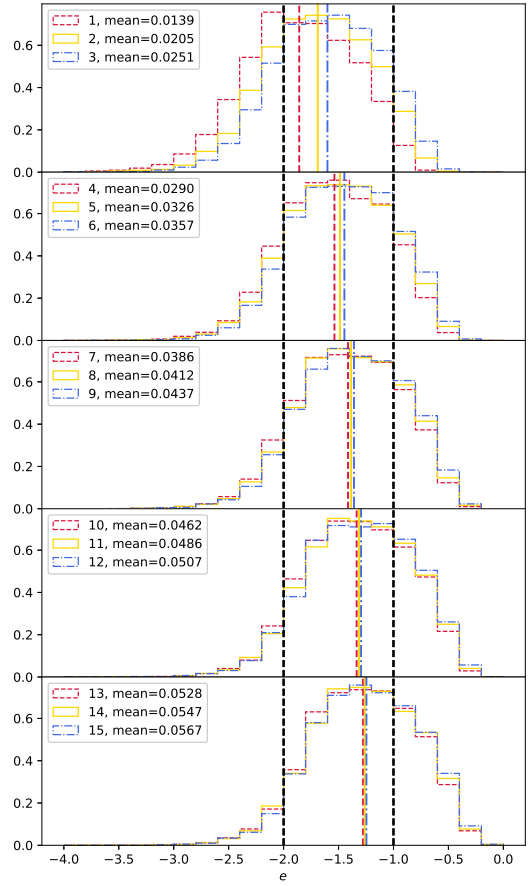}
	\caption{The Probability Distribution Function (PDF) of the eccentricities after 1-15 kicks. The vertical lines with the same color as the PDF lines represent the mean value of the specific population, and the two black dashed vertical lines mark $e=0.01$ and $0.1$ for reference.
	}\label{fig:Fig4}
\end{figure}

\begin{figure}
	\centering
	\includegraphics[width=0.7\textwidth]{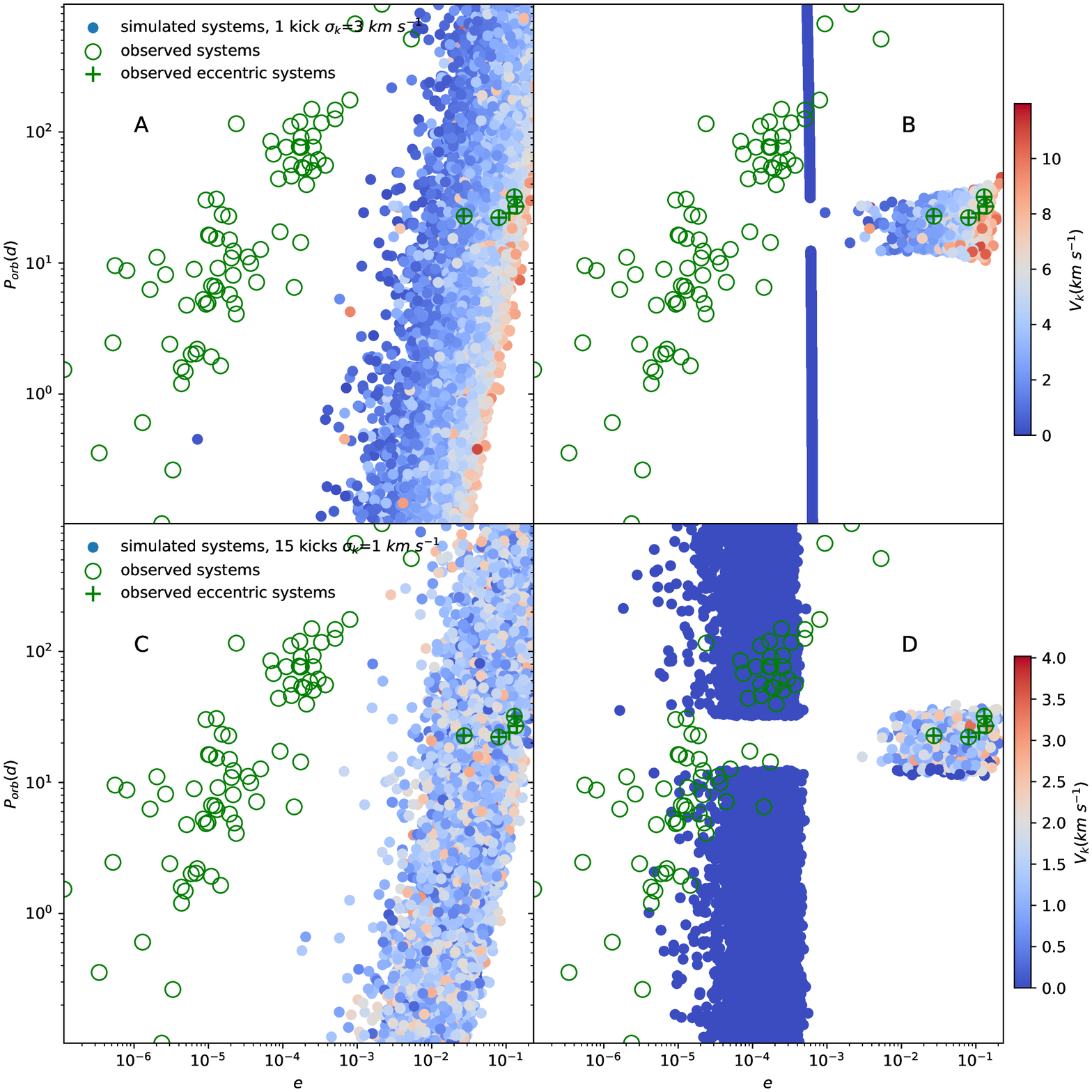}
	\caption{Distributions of the simulated population on the $P_{\rm orb}-e$ plane, overlaid with the observed MSP-He WD binaries. The MSP data are taken from \citet{2018ApJ...864...30H} and the ATNF pulsar catalog \citep{2005AJ....129.1993M}.
	The upper and lower panels correspond to populations $K_{1}\sigma_{3}$ and $K_{15}\sigma_{1}$, respectively. In the left and right panels, all proto-WDs and only those within the mass interval ($[0.268,\ 0.281]\,M_\sun$) are assumed to be subject to kicks, respectively. The grey dashed horizontal lines mark the ``period interval'' translated from the mass interval with the TS99 relation.
	}\label{fig:Fig5}
\end{figure}

\begin{figure}
	\centering
	\includegraphics[width=0.7\textwidth]{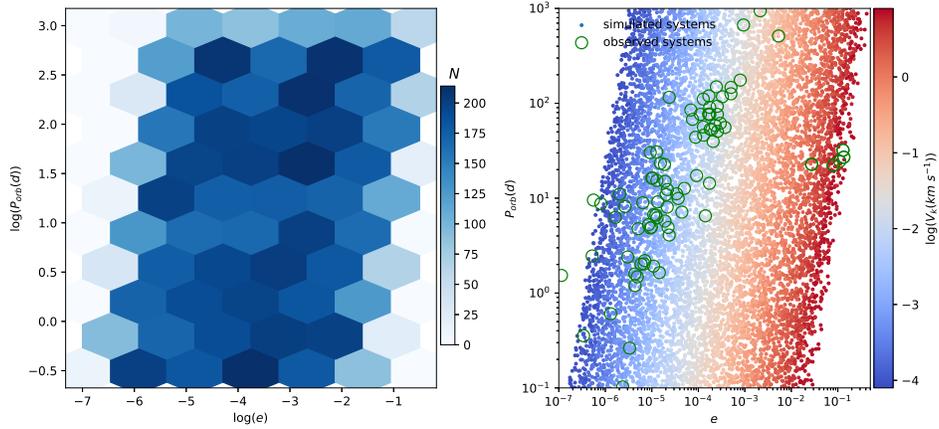}
	\caption{The left panel shows the number distribution of the simulated population on the $P_{\rm orb}-e$ plane, and the right panel compares the simulated and observed distributions on the $P_{\rm orb}-e$ plane. We assume that all proto-WDs experience a kick during the shell flash, and the ejecta mass and kick velocity obey a power-law distribution.
	}\label{fig:Fig6}
\end{figure}

\clearpage

\clearpage
\begin{table}[!h]
\caption{Physical parameters of the five observed eccentric MSP binaries}
\begin{tabular}{lclclclclclcl}
\hline
PSR name &J1950$+$0327 &J2234$+$0611 &J1946$+$3417 &J1618$-$3921 &J0955$-$6150 &\\
binary orbital period $P_{\rm orb}$ (d) & 22.2 & 32.0   & 22.2   & 22.7   & 24.6   &\\
the WD mass $M_{\rm WD}$ ($M_\odot$)   & 0.28  & 0.30  & 0.31    & 0.20($M_{\rm med}$)   & 0.25($M_{\rm med}$)  &\\
the pulsar mass $M_{\rm PSR}$ ($M_\odot$)   & 1.50  & 1.38  & 1.78    & $-$    & $-$   &\\
eccentricity $e$    &0.0798   &0.129  & 0.134   &  0.0274  & 0.110    & \\
spin period, $P$(ms) &4.30   &3.58& 3.17 &12.0  & 2.00   & \\
first derivative of $P$, $\dot{P}((10^{-20} {\rm ss^{-1}}))$ &1.88  &1.20    & 0.314  &5.41&$-$  & \\
transverse velocity $V_{T}( {\rm km\ s^{-1}})$ & $-$  & $145$ & $200 \pm 60$  &$80$, $160$   &$-$   & \\
\hline
Refs. &1-2 &3-5&6-7& 8-10 & 11 & \\
\hline
\end{tabular}

References: 1. \citet{2015ApJ...806..140K}, 2. \cite{2019ApJ...881..165Z},
3. \citet{2013ApJ...775...51D}, 4. \cite{2016ApJ...830...36A}, 5. \cite{2019ApJ...870...74S}, 6. \citet{2013MNRAS.435.2234B}, 7. \cite{2017MNRAS.465.1711B}, 8. \citet{2001ApJ...553..801E}, 9. \cite{2010NewAR..54...80B}, 10. \cite{2018A&A...612A..78O}, 11. \citet{2015ApJ...810...85C}.
\label{table:2}
\end{table}

\end{document}